\documentclass[a4paper]{article}
\usepackage{INTERSPEECH2021}
\usepackage{amsmath,graphicx, multirow,url}
\usepackage{color,soul}
\usepackage{hyperref}
\usepackage{mathtools}

\DeclarePairedDelimiter\floor{\lfloor}{\rfloor}

\title{Citrinet: Closing the Gap between Non-Autoregressive and Autoregressive End-to-End Models for Automatic Speech Recognition}

\name{\begin{tabular}{c} Somshubra Majumdar, Jagadeesh Balam, Oleksii Hrinchuk,\\
Vitaly Lavrukhin, Vahid Noroozi, 
Boris Ginsburg
\thanks{Preprint. Submitted to INTERSPEECH-21}
\end{tabular}
}
\address{NVIDIA, USA}
\email{\{smajumdar,jbalam,ohrinchuk,vlavrukhin,vnoroozi,bginsburg\}@nvidia.com} 

\begin{document}
\maketitle

\begin{abstract}
    We propose Citrinet - a new end-to-end convolutional  Connectionist Temporal Classification (CTC) based automatic speech recognition (ASR) model. Citrinet is  deep residual neural model which uses 1D time-channel separable convolutions combined with sub-word encoding and squeeze-and-excitation.  The resulting architecture significantly reduces the gap between non-autoregressive and  sequence-to-sequence and transducer models. We evaluate Citrinet on  LibriSpeech, TED-LIUM 2,  {AISHELL-1} and Multilingual LibriSpeech (MLS) English speech datasets. Citrinet accuracy on these datasets is close to the best autoregressive Transducer models. 
\end{abstract}


\section{Introduction}
\label{sec:intro}
End-to-end neural ASR models can be roughly classified into three groups based on the network architecture and loss type:\footnote{
See \cite{Battenberg2017, prabhavalkar2017comparison} for a comprehensive comparison between CTC, Seq2Seq, and RNN-T models.
}
\begin{enumerate}
\item
\textit{Connectionist Temporal Classification (CTC)} models \cite{graves2006} 
\item
\textit{Sequence-to-sequence} (Seq2Seq) models with attention, e.g. Listen-Attend-Spell \cite{Chorowski2014,chan2015,Bahdanau2016} or Transformer with Seq2Seq loss \cite{Synnaeve2019}
\item
\textit{RNN-Transducers} (RNN-T)  \cite{graves2012transducer}
\end{enumerate}
Seq2Seq and RNN-T models are autoregressive and  sequential decoding  makes them slower to train and evaluate than non-autoregressive  models. CTC models have the benefit of being more stable and easier to train than autoregressive models, but the latest Seq2Seq models and RNN-Transducers significantly outperform   CTC models, see Table.~\ref{tab:gap}. 
\begin{table}[thb]
\centering
\caption{The accuracy gap between CTC-based models and Seq2Seq and Transducer models, LibriSpeech, WER($\%$)}
\label{tab:gap}
\vspace{4pt}
\scalebox{0.75}{
\begin{tabular}{cccccc} 
 \hline
 \multirow{2}{*}{\textbf{Model}}    &
 \multirow{2}{*}{\textbf{Type}}  &
 \multirow{2}{*}{\textbf{LM}}       &
 \multicolumn{2}{c}{\textbf{Test}}  &
 {\textbf{Params}} \\
  &&& \textbf{clean}& \textbf{other}& M \\
 \hline
 QuartzNet-15x5\cite{kriman2020quartznet} & CTC   & -      & 3.90 & 11.28 & 19 \\
                & & Transf-XL   & 2.69 & 7.25 & \\
\hline
Transformer\cite{Likhomanenko2020} 
  & CTC & - & 2.80 & 7.10 & 270 \\
  & & Transf & 2.10 & 4.70 &  \\
 \hline
 \hline
 LAS+SpecAugm~\cite{park2019} 
   & Seq2Seq  & - & 2.80 & 6.80 & 360 \\
   & & RNN & 2.50 & 5.80 & \\
\hline
Transformer\cite{Synnaeve2019} 
  & Seq2Seq & - & 2.89 & 6.98 & 270 \\
  & & Transf & 2.33 & 5.17 &  \\
\hline
 Conformer-Transf\cite{guo2020espnetconformer} & CTC+Seq2Seq& Transf-XL  & 2.10 & 4.90 & 115 \\
 \hline
 \hline
 Transformer \cite{Zhang2020}
  & Transducer & -  & 2.40 & 5.60 & 139 \\
  & & RNN & 2.00 & 4.60 & \\
\hline
ContextNet-L \cite{han2020contextnet}
  & Transducer & -  & 2.10 & 4.60 & 112\\
  & & RNN & 1.90 & 4.10 & \\
\hline
Conformer-L\cite{gulati2020conformer}
  & Transducer & -  & 2.10 & 4.30 & 118 \\
  & & RNN & 1.90 & 3.90 & \\
 \hline
\end{tabular}
}
\end{table}
The difference in the accuracy between CTC and autoregressive models is usually explained by the claim that 
``because of the strong conditional independence assumption,...CTC does not implicitly learn a language model over the data (unlike the attention-based encoder-decoder architectures). It is therefore \textit{essential} when using CTC to interpolate a language model"~\cite{slp3}. We show that CTC models can overcome these limitations by using recent advances in NN architectures.

In this paper, we describe Citrinet - a  deep convolutional  CTC model.
The Citrinet encoder combines 1D time-channel separable convolutions from QuartzNet \cite{kriman2020quartznet} with the \textit{squeeze-and-excite} (SE)  mechanism \cite{hu2018squeeze} from the ContextNet~\cite{han2020contextnet}. 
Citrinet significantly closes the gap between CTC and the best Seq2Seq and Transducers models. Without any external LM, the Citrinet-1024 model reaches Word Error Rate (WER)
$6.22\%$ on LibriSpeech\cite{panayotov2015librispeech} test-other, 
WER $8.46\%$ on Multilingual LibriSpeech (MLS) English ~\cite{MLS},
WER $8.9\%$ on TED-LIUM2~\cite{tedlium2}, 
and WER $6.8\%$ on  AISHELL-1~\cite{aishell2017} test sets.

\section{Model architecture}

\begin{figure}[t]
 \centering
 \includegraphics[width=1.00\linewidth]{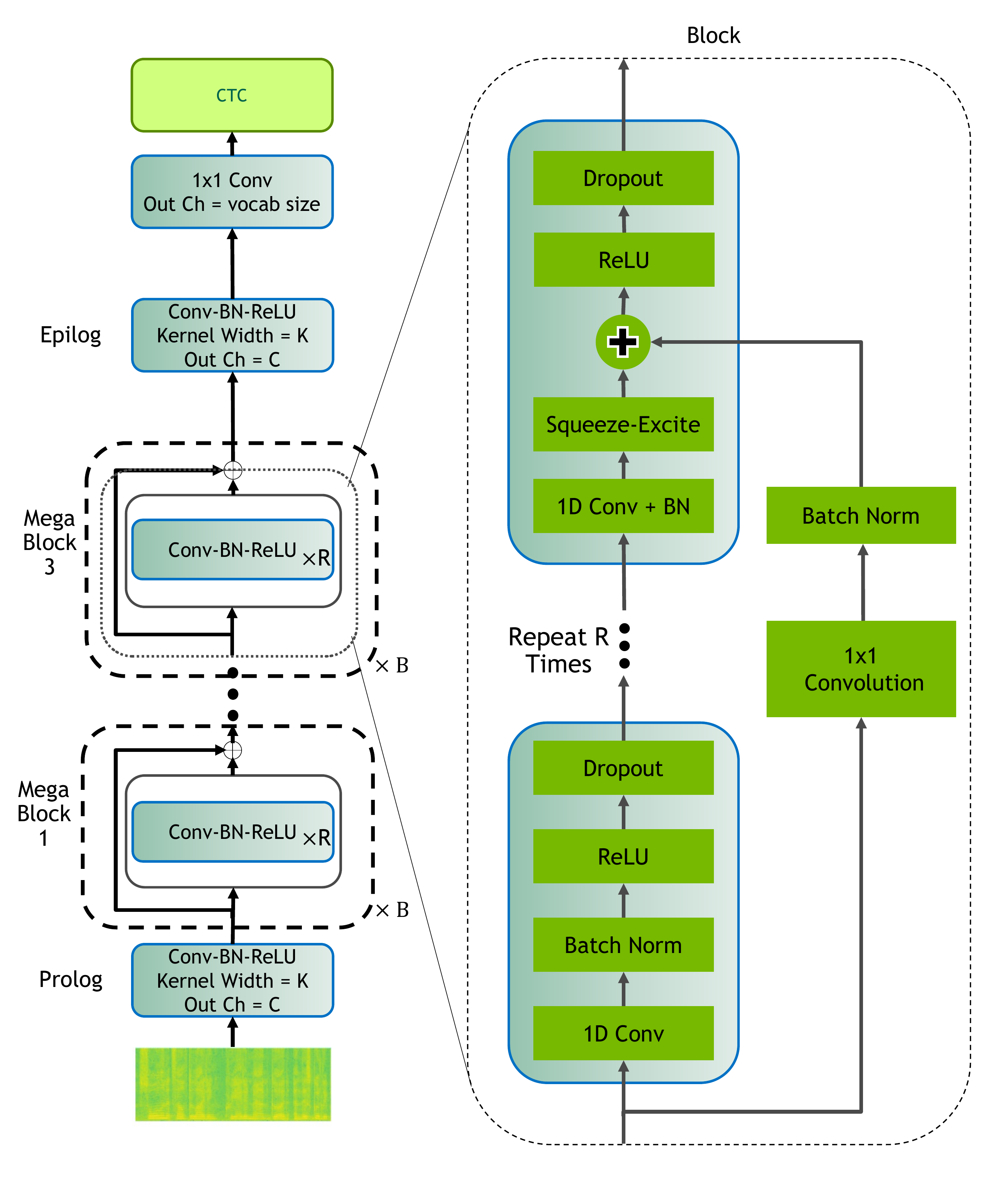}
 \caption{Citrinet BxRxC architecture. Citrinet starts with prolog layer $B_0$, then 3 mega-blocks combined from residual blocks $B_1...B_6$, $B_7...B_{13}$, $B_{14}...B_{21}$, and epilog layer $B_{22}$. Each mega-block begins with a convolutional layer with stride 2. A residual block consists of $R$ ``basic" blocks plus  squeeze-excite module in the end. Basic block is combined from 1D time-channel separable convolution with kernel $K$, batch-norm, ReLU and dropout. All convolutional layers have the same number of channels $C$ except the epilog layer.}
 \label{fig:quartz_arch}
\end{figure}

Citrinet is a 1D time-channel separable convolutional CTC model with a QuartzNet~\cite{kriman2020quartznet} like architecture 
enhanced with 1D Squeeze-and-Excitation (SE) \cite{hu2018squeeze} context modules. 
Fig.~\ref{fig:quartz_arch} describes the Citrinet-$B$x$R$x$C$ model, where $B$ is the number of blocks, $R$ is the number of repeated sub-blocks per block, and $C$ is the number of filters in the convolution layers of each block.  
Citrinet uses the standard acoustic front-end: 80-dimensional log-mel filter banks with a $25$ms window and a stride of $10$ms. 

The network starts with a prolog block $B_0$, then three mega-blocks $B_1\dots B_6, B_7\dots B_{13}, B_{14}\dots B_{21}$, and an epilog module $B_{22}$. 
Each mega-block begins with a 1D time-channel separable convolutional layer with stride 2, so Citrinet progressively down-samples the input three times in the time domain.
A mega-block is combined from residual blocks $B_i$. Each residual block consists of basic QuartzNet blocks, repeated $R$ times, plus an SE module in the end. A QuartzNet block is composed of 1D time-channel separable convolution with kernel $K$, batch-norm, ReLU, and dropout layers. All convolutional layers have the same number of channels $C$ except the epilog.
Citrinet  supports a range of \textit{kernel layouts} for 1D convolutional layers shown in Table~\ref{tab:Kernel_layout}.
\begin{table}[ht!]
\caption{Citrinet: kernel layout configurations. Numbers below are kernel size for 1D convolutional layers.}
\label{tab:Kernel_layout}
\centering
\scalebox{0.68}
{
\begin{tabular}{cccccc}
 \hline
   \textbf{W} &
   $\mathbf{B_0}$ &
   $\mathbf{B_1-B_{6}}$ & 
   $\mathbf{B_7-B_{13}}$ & 
   $\mathbf{B_{14}-B_{21}}$ &
   $\mathbf{B_{22}}$\\
 \hline
  $K_1$ &
    5 &
    3,3,3,5,5,5 & 
    3,3,5,5,5,5,7 &  
    7,7,7,7,9,9,9,9 &
    41\\
  $K_2$ &
    5 &
    5,7,7,9,9,11 &
    7,7,9,9,11,11, 13 &
    13,13,15,15,17,17,19,19 &
    41\\
  $K_3$ & 
   5 &
   9,9,11,13,15,15   &  9,11,13,15,15,17,19 &  19,21,21,23,25,27,27,29 &
   41\\
  \hline
  $K_4$ & 
   5 &
   11,13,15,17,19,21   &  13,15,17,19,21,23,25 &  25,27,29,31,33,35,37,39 &
   41\\
  \hline
\end{tabular}
}
\end{table}
Section 4.1 explains how these kernel layouts were derived. Narrow $K_1$ layout is better for streaming, while wider  layouts are designed for offline mode.
We use $K_4$ as a baseline  throughout the paper.

Citrinet uses the Squeeze-and-Excitation (SE) mechanism from \cite{han2020contextnet} to extend model context. SE block computes a learned global scaling factor per channel  $\theta(x)$
\begin{align*}
    &\theta(x) = \sigma(W_2(ReLU(W_1 \bar{x} + b_1)) + b_2) \text{,} 
\end{align*}
where $W_1, W_2, b_1$ and $b_2$ are trainable weights, $\sigma$ is the sigmoid,  and $\bar{x} = \frac{1}{T} \sum_t {x_t} \text{,}$. The scaling factor $\theta(x)$ is applied via pointwise multiplication to each of the input channels:
\begin{align*}
  & SE(x) = \theta(x) \odot x
\end{align*}

\section{Experiments}


\subsection{LibriSpeech}
We trained five configurations of Citrinet with $K_4$ kernel layout from Table ~\ref{tab:Kernel_layout} and $R=5$  which differed only in the number of channels $C$, on LibriSpeech (LS) dataset~\cite{panayotov2015librispeech}. 
We used a word-piece tokenizer with 256 sub-word units built on the LS train set using the Huggingface library~\cite{wolf2020huggingface}. 
The same tokenizer is used also for two external language models (6-gram and Transformer) for prefix beam re-scoring. Both LMs are constructed using the default LM corpus combined  with text from the LS train transcripts. 
To train Citrinet models we used the NovoGrad optimizer \cite{novograd2019} with learning rate (LR) of $0.05$, $\beta_1=0.8, \beta_2=0.25$ and weight decay of $0.001$.
We used a cosine annealing LR  policy with 1K step linear  warmup. We used  SpecAugment~\cite{park2019} with two frequency masks (F=27), two time masks for small models (C=256, 384) and ten time masks for other models with $5\%$ adaptive mask length. The models have been trained for $1000$ epochs on 32 V100 GPUs with a batch size of $32$ per GPU. 

The results  are shown in Table~\ref{tab:LibriSpeech}.
We use for comparison two Transducers: ContextNet and Conformer. The largest Citrinet model has greedy WER $6.2\%$ test-other set.
With LM rescoring, the gap between Citrinet and SOTA Transducers for test-other is reduced to only $0.59-0.79\%$.
\begin{table}[thb!]
\centering
\caption{LibriSpeech: Citrinet vs Transducers, WER($\%$)}
\vspace{4pt}
\label{tab:LibriSpeech}
\scalebox{1.0}{
\begin{tabular}{cccccc} 
 \toprule
 \multirow{2}{*}{\textbf{Model}}    &
 \multirow{2}{*}{\textbf{LM}}       &
 \multicolumn{2}{c}{\textbf{Test}}  &
 \textbf{Params,}  \\
  &&\textbf{clean}& \textbf{other}& \textbf{M}  \\
\hline
 ContextNet-L \cite{han2020contextnet}
  &  -  & 2.10 & 4.60 & 112.7 \\
  &  RNN & 1.90 & 4.10 & \\
\hline
  Conformer-L\cite{gulati2020conformer}
   & -  & 2.10 & 4.30 & 118 \\
   & RNN & 1.90 & 3.90 & \\
 \hline 
 \hline 
 \multirow{3}{*}{Citrinet-256} &  -      & 3.78 & 9.6 & \multirow{3}{*}{9.8} \\
    & 6-gram  & 3.65 & 8.06 \\
    & Transf  & 2.75 & 6.87 \\
 \midrule
  \multirow{3}{*}{Citrinet-384} 
    & -      & 3.20  & 7.90  &  \multirow{3}{*}{21.0} \\
    & 6-gram & 2.94 & 6.71 & \\
    &  Transf & 2.52 & 5.95 & \\
 \midrule
  \multirow{3}{*}{Citrinet-512}
    &  -     & 3.11 & 7.82 & \multirow{3}{*}{36.5} \\
    & 6-gram & 2.40  & 6.08 & \\
    & Transf & 2.19 & 5.5  & \\
 \midrule
  \multirow{3}{*}{Citrinet-768} 
    & -      & 2.57 & 6.35 & \multirow{3}{*}{81} \\
    & 6-gram & 2.15 & 5.11 & \\
    & Transf & 2.04 & 4.79 & \\
 \midrule
  \multirow{3}{*}{Citrinet-1024}
    & -      & 2.52 & 6.22 & \multirow{3}{*}{142} \\
    & 6-gram & 2.10  & 5.06 & \\
    & Transf & 2.00  & 4.69 & \\
 \bottomrule
\end{tabular}
}
\end{table}



\subsection{MLS}

We trained Citrinet-1024 on the English part of MLS dataset~\cite{MLS}. The original MLS-English train set has 44.66k hours. We filtered out utterances with misaligned transcripts, typos from optical character recognition and ended up with 42.97k hours. To make the model compatible with the LibriSpeech alphabet, we replaced all hyphens with a space character in all splits (train, dev, test). Additionally, we dropped six utterances with Spanish characters from the dev set and one utterance with Russian characters from the test set. Finally, we deleted three instances of the period character from the test set. 
We constructed a 1024 sentence piece tokenizer~\cite{kudo2018sentencepiece} using the  text corpus of the MLS train set. 

We trained the model using the NovoGrad optimizer with a peak LR of 0.005. All other hyper-parameters are the same as for LibriSpeech. The model was trained for $100$ epochs using $256$ GPUs with a batch of 32 per GPU.
Table~\ref{tab:MLS} shows WERs of Citrinet-1024 trained on MLS English and evaluated on LibriSpeech and MLS. 
\begin{table}[thb]
\centering
\caption{MLS English: Citrinet trained on MLS English, evaluated on LibriSpeech-other and on MLS,  WER($\%$)}
\vspace{4pt}
\label{tab:MLS}
\scalebox{1.00}{
\begin{tabular}{ccccccc} 
 \toprule
 \multirow{2}{*}{\textbf{Model}}  &
 \multirow{2}{*}{\textbf{LM}}  &
 \multicolumn{2}{c}{\textbf{LS-other}}  &
 \multicolumn{2}{c}{\textbf{MLS}} 
 \\
  &  & 
  \textbf{dev} & \textbf{test}  & \textbf{dev} & \textbf{test} &
  \\
 \midrule
  & -     & 5.79 & 5.69 & 6.99 & 8.46\\
  Citrinet-1024 
  & 6-gram & 4.72 & 4.83 & 5.76 & 6.79\\
  & Transf & 4.41 & 4.62 & 5.44 & 6.39\\
 \bottomrule
\end{tabular}
}
\end{table}

\subsection{TED-LIUM 2}
\label{sec:tedlium}
We trained two Citrinet-1024 models for the TED-LIUM 2 corpus \cite{tedlium2} with 207 hours of speech.
The first model was trained from scratch (TS), and the second model was fine-tuned (FT) from the model pre-trained on MLS. We used 1024 sentence piece tokens trained on the training set of the TED-LIUM 2 for the TS model and used the tokenizer trained on MLS data for the FT model. For the LM (N-grams and Transformer), we used both TED-LIUM 2 text for LM and the texts from the train part.
The TS model was trained for 1000 epochs using the NovoGrad optimizer with a peak LR of 0.05. Fine-tuning was done for 200 epochs using  with a peak LR of 0.005. Both models were trained on 16 GPUs with a batch size of 32 per GPU. All other hyper-parameters are the same as the LibriSpeech model.
For comparison we used two models: RWTH hybrid HMM-based model trained with SpecAugment \cite{zhou2020rwth}, and end-to-end model composed from 7-layer time-delay NN combined with 3 LSTM layers  \cite{CAPIO2017}. 

Table~\ref{tab:tedlium} shows the Citrinet-1024 evaluation on TED-LIUM 2 \textit{dev} and \textit{test} sets.
The Citrinet-1024 fine-tuned from the  pre-trained  on MLS model matches the hybrid HMM model~\cite{zhou2020rwth} and sets a new SOTA for end-to-end NN-based models.
\begin{table}[thb]
\centering
\caption{TED-LIUM rev.2:  Citrinet-1024 trained from scratch  (TS) and fine-tuned (FT) from MLS English, WER($\%$)}
\vspace{4pt}
\label{tab:tedlium}
\scalebox{1.00}{
\begin{tabular}{ccccc} 
 \hline
 \textbf{Model} &  \textbf{LM} &
 \textbf{dev}   & \textbf{test} \\
 \hline
 Dense TDNN-LSTM \cite{CAPIO2017}
    & N-gram & 7.60 & 8.10 \\
    & RNN    & 7.10 & 7.70 \\
 \hline
 RWTH Hybrid HMM \cite{zhou2020rwth}
    & 4-gram & 6.80 & 7.30 \\
    & Transf & 5.10 & 5.60 \\
 \hline
 \hline
  Citrinet-1024/TS
    & - & 9.79 & 8.90 \\
    & 6-gram & 7.75 &  7.47 \\
 \hline
  Citrinet-1024/FT
   & -      & 6.69 & 6.05 \\
   & 6-gram & 5.89 & 5.65\\
\hline
\end{tabular}
}
\end{table}


\subsection{AISHELL-1}
\label{sec:aishell}
We trained two  Citrinet-1024 models for  AISHELL-1 \cite{aishell2017}, a Mandarin corpus with 150 hours of training speech. The first model was trained from scratch (TS). 
For the second model (FT) weights of the encoder were pre-initialized  from the encoder of the English model trained on MLS. Unlike the English language models, we used character level tokenization instead of sentence piece tokenization for Mandarin ASR. 
The {AISHELL-1} TS and FT models were trained with the same parameters as the TED-LIUM 2 models from Section~\ref{sec:tedlium}.

Table~\ref{tab:aishell} compares Character Error Rate (CER) for Citrinet and two hybrid CTC-attention models:
ESPnet Transformer~\cite{Karita2019} and U2 model \cite{zhang2020unified} composed from Conformer encoder and two decoders - CTC  and Transformer. 
Citrinet-1024 fine-tuned from the English model, has better CER then Transformer-based Seq2Seq model\cite{Karita2019}.
\begin{table}[thb]
\centering
\caption{AISHELL-1: Citrinet-1024 trained from scratch (TS) and Citrinet-1024 fine-tuned (FT) from the model pretrained on MLS English, CER($\%$)}
\vspace{4pt}
\label{tab:aishell}
\scalebox{1.00}{
\begin{tabular}{cccc} 
 \hline
 \textbf{Model} &  \textbf{LM} &  \textbf{dev} & \textbf{test} \\
 \hline
 ESPNet Transformer\cite{Karita2019} &   &6.00 & 6.70 \\
 \hline
 U2 Conformer-Transformer \cite{zhang2020unified}
    &- & - & 4.72 \\
 \hline
 \hline
  Citrinet-1024/TS
    & - & 6.16 & 6.82 \\
    & 4-gram & 5.89 & 6.39 \\
 \hline
  Citrinet-1024/FT
    & -      & 5.2 & 5.71  \\
    & 4-gram & 5.2 & 5.55 \\
\hline
\end{tabular}
}
\end{table}

\section{Ablation study}

We perform an ablation study to determine the contribution of each component to the model accuracy. We use Citrinet-$B$x$R$x$C$ with $B=21$, $R=5$, $C=384$ and kernel layout $K_4$ as baseline. For brevity, we denote the configuration of Citrinet-$B$x$R$x$C$ as Citrinet-$C$, such that it expands to Citrinet-21x5x$C$. As a baseline we use 1024 Word Piece Tokens for the tokenizer vocabulary. 
The ablation study is done  on the LibriSpeech dataset. All models have been trained for 400 epochs on 32 GPUs with a batch of 32 per GPU. We used the NovoGrad optimizer with betas of $(0.8, 0.25)$, cosine learning rate decay with peak LR of 0.05, and weight decay of $0.001$.

\subsection{Kernel width}
\label{Ablation_kernel}

We use as a baseline 
the wide $K_4$ kernel layout in Table~\ref{tab:Ablation_Kernel}.
\begin{table}[ht!]
\caption{Citrinet: baseline kernel layout.}
\label{tab:Ablation_Kernel}
\centering
\scalebox{1.00}
{
\begin{tabular}{c c }
 \hline
   \textbf{Block} &  \textbf{Filters, K} \\
 \hline
 $B_0$ & 5 \\
 $B_1-B_6$       & 11,13,15,17,19,21\\
 $B_7-B_{13}$    & 13,15,17,19,21,23,25\\
 $B_{14}-B_{21}$ & 25,27,29,31,33,35,37,39\\
 $B_{22}$ & 41 \\
  \hline
\end{tabular}
}
\end{table}

We scale all kernels except  the prologue and epilogue layers with the same scaling factor, $\gamma$: the scaled kernel widths are  $k'_{i} = \floor{k_{i} * \gamma}$, where $k_i$ is the original kernel width for the $i$-th block. If $k'_{i}$ width is even, it is incremented by 1.
Table~\ref{tab:Ablation_kernels_size} shows the model WER for  $\gamma=0.25,0.5,0.75,1$. 
\begin{table}[thb]
\centering
\caption{Accuracy vs kernel size ($\gamma$ - kernels scaling factor). Citrinet-384, LibriSpeech, greedy WER($\%$).}
\vspace{4pt}
\label{tab:Ablation_kernels_size}
\scalebox{0.95}{
\begin{tabular}{cccccc} 
 \hline
   \textbf{Kernel}  &
   {\textbf{scaling factor}} &
  \multicolumn{2}{c}{\textbf{dev}} &
  \multicolumn{2}{c}{\textbf{test}} \\
   \textbf{layout}&
   $\mathbf\gamma$ &
   \textbf{clean} & \textbf{other} & \textbf{clean}& \textbf{other} \\
 \hline
  $K_1$ & 0.25 &
  3.36 & 8.89 & 3.56 & 9.28 \\
  $\mathbf{K_2}$ & \textbf{0.50} &
  \textbf{3.26} & \textbf{8.59} & \textbf{3.41} & \textbf{8.74}\\
   $K_3$ & 0.75 &
   3.15 & 8.75 & 3.43 & 9.04 \\
   $K_4$ & 1.00 & 
   3.49 & 9.31 & 3.62 & 9.28 \\
 \hline
\end{tabular}
}
\end{table}

We found that there is an optimal kernel width, and too narrow or too wide kernels lead to higher WER.
The narrow  kernel layout corresponding to $\gamma=0.25$ is a primary candidate for streaming ASR. One can get slightly better accuracy by doubling its receptive field. Increasing the kernel size further will decrease the model accuracy.

\subsection{Scaling network in width and depth}
\label{Ablation_scaling}

We start the scalability study by changing the number of channels $C$ while keeping $R=5$ and the kernel layout as in Table~\ref{tab:Ablation_Kernel}. 
Table~\ref{tab:Ablation_channels} compares the WER and number of parameters for Citrinet with $C=256, 384$ and $512$ channels:
\begin{table}[thb]
\centering
\caption{Accuracy vs model width ($C$ - number of channels). All Citrinet models have R=5. LibriSpeech, greedy WER($\%$).}
\label{tab:Ablation_channels}
\vspace{4pt}
\scalebox{1.00}{
\begin{tabular}{cccccc} 
 \hline
 \multirow{2}{*}{\textbf{ C}}    &
 \multicolumn{2}{c}{\textbf{dev}} &
  \multicolumn{2}{c}{\textbf{test}} & 
   {\textbf{Params,}} \\
  & \textbf{clean} & \textbf{other} & \textbf{clean}& \textbf{other} & \textbf{M}\\
 \hline
 256 & 4.38 & 11.41 & 4.57 & 11.54 & 10.2 \\
 384 & 3.49 & 9.31 & 3.62  & 9.28 & 21.1 \\
 \textbf{512} & \textbf{3.16} & \textbf{8.71} & \textbf{3.27} &\textbf{ 8.83} & \textbf{37.2} \\
 \hline
\end{tabular}
}
\end{table}

Models can be also scaled in depth by changing the number of sub-blocks $R$ per  block. The capacity of the model, its depth and the receptive field of the network grow with the number of repeated blocks, increasing the model accuracy.
Table \ref{tab:Ablation_scaling_depth} compares WER for Citrinet-384 with  various values of $R$. 
\begin{table}[thb!]
\centering
\caption{Accuracy vs model depth 
($R$ - the number of sub-blocks per block). Citrinet-384. LibriSpeech, greedy WER($\%$).}
\vspace{4pt}
\label{tab:Ablation_scaling_depth}
\scalebox{1.00}{
\begin{tabular}{cccccc} 
 \hline
 \multirow{2}{*}{\textbf{R}} &
 \multicolumn{2}{c}{\textbf{dev}} &
  \multicolumn{2}{c}{\textbf{test}} & 
   {\textbf{Params,}} \\
  & \textbf{clean} & \textbf{other} & \textbf{clean}& \textbf{other} & \textbf{M} \\
 \hline
   2 & 4.22 & 11.16 & 4.39 & 11.14 & 11.6 \\
   3 & 3.59 & 10.07 &  3.81 & 9.94 & 14.9 \\
   4 & 3.46 & 9.52 &  3.68 & 9.48 & 18.1 \\
  \textbf{5} & \textbf{3.49} & \textbf{9.31} & \textbf{3.62} & \textbf{9.28} & \textbf{21.1} \\
 \hline
\end{tabular}
}
\end{table}

\subsection{Tokenizer lexicon size}
\label{Ablation_downsampling}

The simple way to reduce the memory usage and increase training and inference speed is to compress intermediate activations along the time dimension. But down-sampling in time is limited by CTC loss which requires the output of the acoustic model to be longer than the target transcription. For example, char-based English CTC models work best at 50 steps per second \cite{Battenberg2017,DeepSpeech2}. 
To bypass this constraint, we use word-piece encoding  \cite{wolf2020huggingface} or byte-pair encoding to represent the input text using fewer encoded tokens. Using sub-word encoding we are able to train models with 8x contraction on the time dimension.
Table \ref{tab:Ablation_lexicon_size} compares the effect of the lexicon size on the model accuracy. 
\begin{table}[thb]
\centering
\caption{Accuracy vs tokenizer lexicon size. Citrinet-384, LibriSpeech, greedy WER ($\%$).}
\vspace{4pt}
\label{tab:Ablation_lexicon_size}
\scalebox{1.00}{
\begin{tabular}{cccccc} 
 \hline
 \multirow{2}{*}{\textbf{Lexicon Size}} &
 \multicolumn{2}{c}{\textbf{dev}} &
  \multicolumn{2}{c}{\textbf{test}} \\
  & \textbf{clean} & \textbf{other} & \textbf{clean}& \textbf{other} \\
 \hline
  128  & 3.50 & 9.64 & 3.63 & 9.42 \\
  \textbf{256} &\textbf{3.30} &
  \textbf{8.88} &\textbf{ 3.46} & \textbf{8.92} \\
  512  & 3.35 & 9.53 & 3.77 & 9.52 \\
  1024 & 3.49 & 9.31 & 3.62 & 9.28 \\
  2048  & 3.81 & 9.99 & 4.11 & 10.59 \\
  4096  & 7.09 & 14.73 & 7.18 & 14.44\\
 \hline
\end{tabular}
}
\end{table}

The very large lexicon sizes tend to significantly harm transcription accuracy. On the other side, we observed that if the lexicon size is less than 128, CTC loss cannot be computed on a significant number of transcripts that exceed the length of the output of the acoustic model after 8x compression in the time domain. 

\subsection{Context window size}
\label{Ablation_context_size}

To study the effect of the context window size of the SE module, we used Citrinet-512 with 1024 Word Piece tokenization.
We replace the global pooling operator with an average pooling operator where the pooling size defines the local context windows of 256, 512, and 1024.
The results are shown in the Table~ \ref{tab:Ablation_context_window_study}.
The context provided by the SE mechanism significantly contributes to the Citrinet accuracy, similar to   \cite{han2020contextnet}. 
\begin{table}[thb]
\centering
\caption{Accuracy vs squeeze-excite context window size. Citrinet-512, LibriSpeech, greedy WER($\%$).}
\vspace{4pt}
\label{tab:Ablation_context_window_study}
\scalebox{1.00}{
\begin{tabular}{c c c c c} 
 \hline
    \multirow{2}{*}{\textbf{SE window}}    &
     \multicolumn{2}{c}{\textbf{dev}} &
  \multicolumn{2}{c}{\textbf{test}} \\
  & \textbf{clean} & \textbf{other} & \textbf{clean}& \textbf{other} \\
 \hline
 -    & 3.39 & 9.05  & 3.56 & 9.09 \\
 256  & 3.01 & 8.59  & 3.36 & 8.61 \\
 512  & 2.96 & 8.12  & 3.27 & 8.26 \\
 1024 & 2.91 & 8.03  & 3.15 & 8.04 \\
 \textbf{Global} & \textbf{2.86} & \textbf{8.02}  & \textbf{3.12} & \textbf{7.99} \\
 \hline
\end{tabular}
}
\end{table}

\section{Conclusions}
\label{sec:conclusions}
In this paper, we introduced Citrinet - a new end-to-end non-autoregressive CTC-based model. Citrinet enhances the QuartzNet \cite{kriman2020quartznet} architecture with the Squeeze-and-Excitation mechanism from ContextNet \cite{han2020contextnet}. 
Citrinet significantly reduces the gap between non-autoregressive and state-of-the-art autoregressive Seq2Seq and RNN-T models \cite{Battenberg2017, prabhavalkar2017comparison}. 
Contrary to the common belief that ``\textit{CTC requires an external language model to output meaningful results}''\cite{deepspeech3} 
, Citrinet models demonstrate very high accuracy without any external language model on LibriSpeech, MLS, TED-LIUM 2,  and AI-SHELL datasets.

The models and training recipes have been released in the NeMo toolkit~\cite{kuchaiev2019nemo}.\footnote{\url{https://github.com/NVIDIA/NeMo}} 

\section{Acknowledgments}
The authors thank the NVIDIA AI Applications team for the helpful feedback and review.

\bibliography{refs}

\begin{thebibliography}{10}
\providecommand{\url}[1]{#1}
\csname url@samestyle\endcsname
\providecommand{\newblock}{\relax}
\providecommand{\bibinfo}[2]{#2}
\providecommand{\BIBentrySTDinterwordspacing}{\spaceskip=0pt\relax}
\providecommand{\BIBentryALTinterwordstretchfactor}{4}
\providecommand{\BIBentryALTinterwordspacing}{\spaceskip=\fontdimen2\font plus
\BIBentryALTinterwordstretchfactor\fontdimen3\font minus
  \fontdimen4\font\relax}
\providecommand{\BIBforeignlanguage}[2]{{%
\expandafter\ifx\csname l@#1\endcsname\relax
\typeout{** WARNING: IEEEtran.bst: No hyphenation pattern has been}%
\typeout{** loaded for the language `#1'. Using the pattern for}%
\typeout{** the default language instead.}%
\else
\language=\csname l@#1\endcsname
\fi
#2}}
\providecommand{\BIBdecl}{\relax}
\BIBdecl

\bibitem{Battenberg2017}
E.~{Battenberg}, J.~{Chen}, R.~{Child}, A.~{Coates}, Y.~G.~Y. {Li}, H.~{Liu},
  S.~{Satheesh}, A.~{Sriram}, and Z.~{Zhu}, ``Exploring neural transducers for
  end-to-end speech recognition,'' in \emph{ASRU}, 2017.

\bibitem{prabhavalkar2017comparison}
R.~Prabhavalkar, K.~Rao, T.~N. Sainath, B.~Li, L.~Johnson, and N.~Jaitly, ``A
  comparison of sequence-to-sequence models for speech recognition,'' in
  \emph{Interspeech}, 2017.

\bibitem{graves2006}
A.~Graves, S.~Fern{\'a}ndez, F.~Gomez, and J.~Schmidhuber, ``Connectionist
  temporal classification: labelling unsegmented sequence data with recurrent
  neural networks,'' in \emph{ICML}, 2006.

\bibitem{Chorowski2014}
J.~Chorowski, D.~Bahdanau, K.~Cho, and Y.~Bengio, ``End-to-end continuous
  speech recognition using attention-based {Recurrent NN}: First results,'' in
  \emph{NIPS, Deep Learning and Representation Learning Workshop}, 2014.

\bibitem{chan2015}
W.~Chan, N.~Jaitly, Q.~V. Le, and O.~Vinyals, ``Listen, attend and spell,'' in
  \emph{ICASSP}, 2016.

\bibitem{Bahdanau2016}
D.~Bahdanau, J.~Chorowski, D.~Serdyuk, P.~Brakel, and Y.~Bengio, ``End-to-end
  attention-based large vocabulary speech recognition.'' in \emph{ICASSP},
  2016.

\bibitem{Synnaeve2019}
G.~Synnaeve, Q.~Xu, J.~Kahn, E.~Grave, T.~Likhomanenko, V.~Pratap, A.~Sriram,
  V.~Liptchinsky, and R.~Collobert, ``End-to-end {ASR}: from supervised to
  semi-supervised learning with modern architectures,''
  \emph{arXiv:1911.08460}, 2019.

\bibitem{graves2012transducer}
A.~Graves, ``Sequence transduction with recurrent neural networks,''
  \emph{arXiv:1211.3711}, 2012.

\bibitem{kriman2020quartznet}
S.~Kriman, S.~Beliaev, B.~Ginsburg, J.~Huang, O.~Kuchaiev, V.~Lavrukhin,
  R.~Leary, J.~Li, and Y.~Zhang, ``{QuartzNet}: Deep automatic speech
  recognition with 1d time-channel separable convolutions,'' in \emph{ICASSP},
  2020.

\bibitem{Likhomanenko2020}
T.~Likhomanenko, Q.~Xu, V.~Pratap, P.~Tomasello, J.~Kahn, G.~Avidov,
  R.~Collobert, and G.~Synnaeve, ``Rethinking evaluation in {ASR}: Are our
  models robust enough?'' \emph{arXiv:2010.11745}, 2020.

\bibitem{park2019}
D.~Park, W.~Chan, Y.~Zhang, C.~Chiu, B.~Zoph, E.~Cubuk, and Q.~Le,
  ``{SpecAugment}: A simple data augmentation method for automatic speech
  recognition,'' \emph{arXiv:1904.08779}, 2019.

\bibitem{guo2020espnetconformer}
P.~Guo, F.~Boyer, X.~Chang, T.~Hayashi, Y.~Higuchi, H.~Inaguma, N.~Kamo, C.~Li,
  D.~Garcia-Romero, J.~Shi \emph{et~al.}, ``Recent developments on {ESPnet}
  toolkit boosted by {Conformer},'' \emph{arXiv:2010.13956}, 2020.

\bibitem{Zhang2020}
Q.~{Zhang}, H.~{Lu}, H.~{Sak}, A.~{Tripathi}, E.~{McDermott}, S.~{Koo}, and
  S.~{Kumar}, ``{Transformer Transducer}: A streamable speech recognition model
  with {Transformer} encoders and {RNN-T} loss,'' in \emph{ICASSP}, 2020.

\bibitem{han2020contextnet}
W.~Han, Z.~Zhang, Y.~Zhang, J.~Yu, C.-C. Chiu, J.~Qin, A.~Gulati, R.~Pang, and
  Y.~Wu, ``Contextnet: Improving convolutional neural networks for automatic
  speech recognition with global context,'' \emph{arXiv:2005.03191}, 2020.

\bibitem{gulati2020conformer}
A.~Gulati, J.~Qin, C.-C. Chiu, N.~Parmar, Y.~Zhang, J.~Yu, W.~Han, S.~Wang,
  Z.~Zhang, Y.~Wu, and R.~Pang, ``Conformer: Convolution-augmented transformer
  for speech recognition,'' \emph{arXiv:2005.08100}, 2020.

\bibitem{slp3}
D.~Jurafsky and J.~H. Martin, \emph{Speech and Language Processing}.\hskip 1em
  plus 0.5em minus 0.4em\relax Preprint, 2020.

\bibitem{hu2018squeeze}
J.~Hu, L.~Shen, and G.~Sun, ``Squeeze-and-excitation networks,'' in
  \emph{ICVPR}, 2018.

\bibitem{panayotov2015librispeech}
V.~Panayotov, G.~Chen, D.~Povey, and S.~Khudanpur, ``Librispeech: an {ASR}
  corpus based on public domain audio books,'' in \emph{ICASSP}, 2015, pp.
  5206--5210.

\bibitem{MLS}
V.~Pratap, Q.~Xu, A.~Sriram, G.~Synnaeve, and R.~Collobert, ``{MLS}: A
  large-scale multilingual dataset for speech research,'' \emph{Interspeech},
  2020.

\bibitem{tedlium2}
A.~Rousseau, P.~Del{\'e}glise, and Y.~Est{\`e}ve, ``Enhancing the {TED}-{LIUM}
  corpus with selected data for language modeling and more {TED} talks,'' in
  \emph{Proceedings of the Ninth International Conference on Language Resources
  and Evaluation ({LREC}'14)}, 2014.

\bibitem{aishell2017}
B.~Hui, D.~Jiayu, N.~Xingyu, W.~Bengu, and H.~Zheng, ``{AIShell}-1: An
  open-source mandarin speech corpus and a speech recognition baseline,'' in
  \emph{Oriental COCOSDA}, 2017.

\bibitem{wolf2020huggingface}
T.~Wolf, L.~Debut, V.~Sanh, J.~Chaumond, C.~Delangue, A.~Moi, P.~Cistac,
  T.~Rault, R.~Louf, M.~Funtowicz, J.~Davison, S.~Shleifer, P.~von Platen,
  C.~Ma, Y.~Jernite, J.~Plu, C.~Xu, T.~L. Scao, S.~Gugger, M.~Drame, Q.~Lhoest,
  and A.~M. Rush, ``{HuggingFace's Transformers}: State-of-the-art natural
  language processing,'' \emph{arXiv:1910.03771}, 2020.

\bibitem{novograd2019}
B.~Ginsburg, P.~Castonguay, O.~Hrinchuk, O.~Kuchaiev, V.~Lavrukhin, R.~Leary,
  J.~Li, H.~Nguyen, and C.~J. M., ``Stochastic gradient methods with layer-wise
  adaptive moments for training of deep networks,'' \emph{arXiv:1905.11286},
  2019.

\bibitem{kudo2018sentencepiece}
T.~Kudo and J.~Richardson, ``{SentencePiece}: A simple and language independent
  subword tokenizer and detokenizer for neural text processing,''
  \emph{arXiv:1808.06226}, 2018.

\bibitem{zhou2020rwth}
W.~Zhou, W.~Michel, K.~Irie, M.~Kitza, R.~Schlüter, and H.~Ney, ``The {RWTH
  ASR System for TED-LIUM Release 2}: Improving hybrid {HMM with
  SpecAugment},'' in \emph{ICASSP}, 2020.

\bibitem{CAPIO2017}
K.~Han, A.~Chandrashekaran, J.~Kim, and I.~R. Lane, ``The {CAPIO} 2017
  conversational speech recognition system,'' \emph{arXiv:1801.00059}, 2018.

\bibitem{Karita2019}
S.~{Karita}, N.~{Chen}, T.~{Hayashi}, T.~{Hori}, H.~{Inaguma}, Z.~{Jiang},
  M.~{Someki}, N.~E.~Y. {Soplin}, R.~{Yamamoto}, X.~{Wang}, S.~{Watanabe},
  T.~{Yoshimura}, and W.~{Zhang}, ``A comparative study on {Transformer vs RNN}
  in speech applications,'' in \emph{ASRU Workshop}, 2019.

\bibitem{zhang2020unified}
B.~Zhang, D.~Wu, Z.~Yao, X.~Wang, F.~Yu, C.~Yang, L.~Guo, Y.~Hu, L.~Xie, and
  X.~Lei, ``Unified streaming and non-streaming two-pass end-to-end model for
  speech recognition,'' \emph{arXiv:2012.05481}, 2020.

\bibitem{DeepSpeech2}
D.~Amodei and etc, ``Deep speech 2: End-to-end speech recognition in english
  and mandarin,'' in \emph{ICML}, 2016.

\bibitem{deepspeech3}
{Baidu Research blog}, ``{Deep Speech 3}: Even more end-to-end speech
  recognition,'' \url{http://research.baidu.com/Blog/index-view?id=90}, 2017.

\bibitem{kuchaiev2019nemo}
O.~Kuchaiev, J.~Li, H.~Nguyen, O.~Hrinchuk, R.~Leary, B.~Ginsburg, S.~Kriman,
  S.~Beliaev, V.~Lavrukhin, J.~Cook \emph{et~al.}, ``{NeMo}: a toolkit for
  building {AI} applications using neural modules,'' \emph{arXiv:1909.09577},
  2019.

\end{thebibliography}
\end{document}